\documentclass{article}
\pdfoutput=1
\usepackage{indentfirst}
\usepackage{amsmath}
\usepackage{graphicx}
\usepackage{hyperref}
\usepackage{amssymb}

\author{Stanislav Srednyak}
\title{Reduction of Feynman integrals to integrals of Schl\"{a}fli functions}
\begin{document}
\maketitle

\abstract{

We show that off-shell perturbative amplitudes with arbitrary number of external lines and complex masses can be reduced to $I$-fold integrals of the generalized Schl\"{a}fli functions, where $I$ is the number of lines in the corresponding vacuum diagram and does not depend on the number of external lines. The Schl\"{a}fli functions are obtained as analytic continuation of the volume of the spherical simplex as a function of the hyperplane parameters that define the simplex. These functions have nice and thoroughly studied analytic, geometric and number theoretic properties. They posess Gauss-Manin connection and in conformal case are expressed by iterated integrals. Our representation sheds new light on geometry of particle configuration spaces in perturbation theory.

\section{Introduction}

In  this note we demonstrate that generic Feynman integral, at generic off-shell values of momenta and different masses, at $L$ loops and arbitrary number of external off-shell lines can be reduced to $I$-fold integral of generalized Schl\"{a}fli function, where is $I$ is the number of lines in the corresponding vacuum diagram. This number is independent of the number of external lines.  The remaining integrations are perforemed over the standard unit simples in $I$ dimensions. The matrix argument of the Schl\"{a}fli function is a rational function constructed from external momenta and certain combinations of Feynman parameters. Schl\"{a}fli functions are functionsdefined on the space of complex $n\times n$ matrices. For real parameters they are basically volumes of the appropriate spherical triangle cut out by a set of hyperplanes. It is remarkable that there is connection between the spaces of particle momenta and basic stratifications of matrix spaces. Our results suggest that some of the results of ~\cite{Nima1} can be generalized to arbitrary theories and off-shell amplitudes. The Schl\"{a}fli function at complex parameters are obtained by analytic continuation. The singularities of Schl\"{a}fli functions are given by vanishing of minors of all possible sizes of the matrix of parameters. Schl\"{a}fli function have good analytic and geometric properties, in particular, it poseses Gauss-Manin connection. The dimensionality of the Schl\"{a}fli function depends on the diagram and dimensionality of space-time.

The motivation for this paper comes from the long-standing problem of finding the overdetermined system of differential equations for Feynman integrals. It has long been a folklore that such systems exist ( "Laporta algorithm") but the equations were established only in a few cases ~\cite{Henn1}. In the mathematics literature such systems are known as holonomic D-modules. If the geometry of the variety of parameters is nice enough, this D-module can be rewritten as a Gauss-Manin connection ~\cite{Manin1}. The relation is not strightforward and D-modules offer more economic way to write combinatorial information. This problem is related to the problem of finding master integrals, as these integrals provide a basis of sections of the holomorphic vector bundle defined by the integral over the generic stratum in the parameter space. These ideas have been successfully applied to Euler integrals in ~\cite{GZK_toric}. It would seem that perturbative integrals in Feynman parametrization are immediately amenable to such analysis. But the dimensionality of the space of Feynman parameters may be large, and direct calculation of the lattice of relations between coordinates of vertices of the Newton polytope is difficult. Therefore it is desirable to reduce the number of Feynman parameters. This is accomplished in the present paper.

The paper is organized as follows. In sec.2 we collect the necessary definitions. In sec. 3 we state and prove the main theorem. In sec. 4 we give 2-loop example. Sec.5 is the conclusion. In appendix we collect several facts concerning the Schl\"{a}fli function.

\section{Preliminaries}

Our starting point is the following expression for the Feynman integral ~\cite{Weinz1}( cartooned at Fig. 1, considered for arbitrary off-shell complex external momenta)
\begin{equation}
I_\omega(p_{i\alpha}=\int_\bigtriangleup \delta(1-\sum x)(dx_{i\alpha})x^\omega F^w G^v
\end{equation}
In this expression, the integral is performed over the unit simplex ( relative chain). $\omega$ is a multi-index. $F,G$ are so-called graph polynomials that are expressed through external momenta and combinatoric parameters of the Feynman graph. We need a few definitions to write them explicitly. 

\begin{figure}
\centering\includegraphics[width=0.5\textwidth]{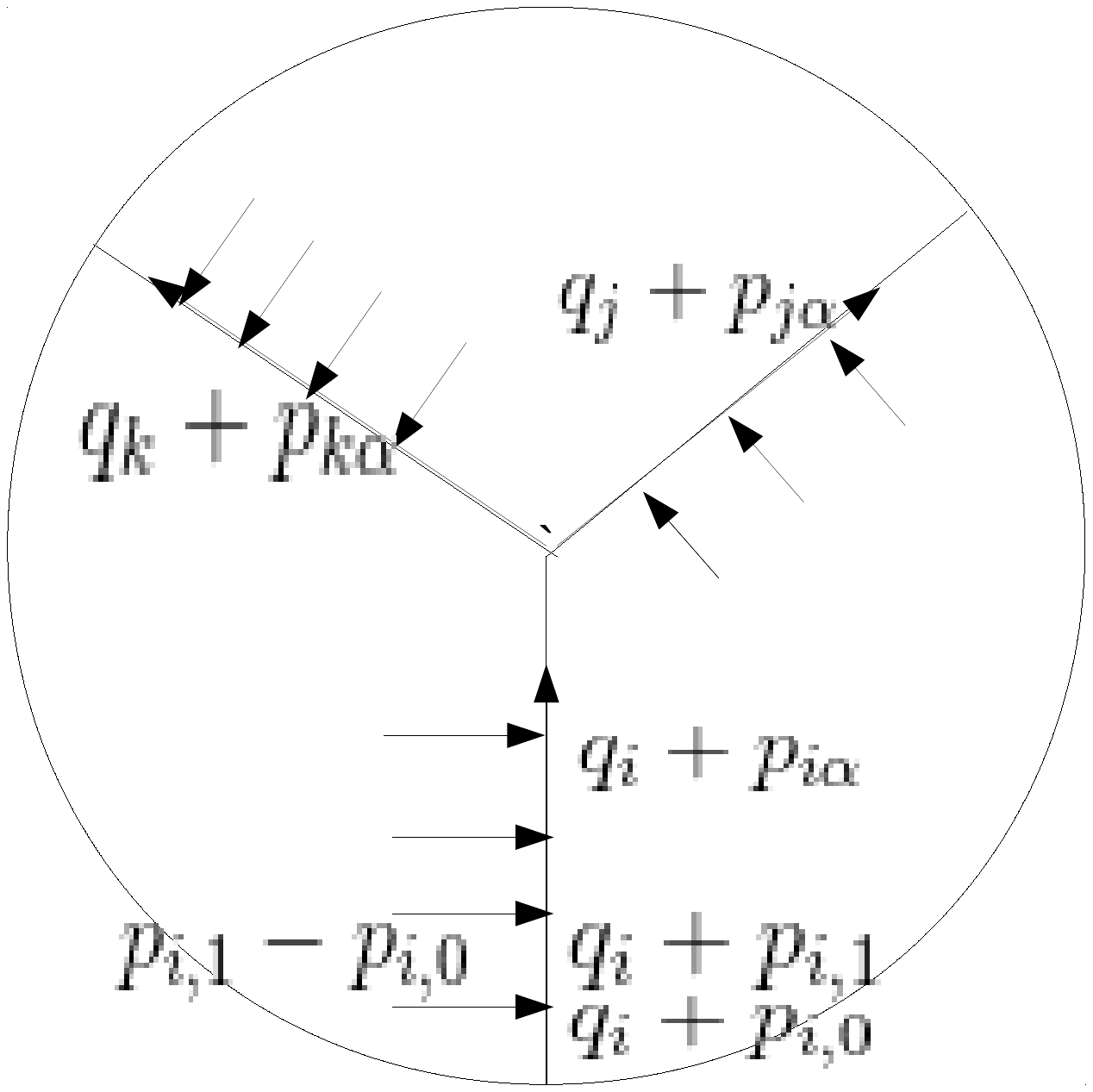}
\caption{A typical diagram containing a triple vertex. Internal lines have two indices $i,\alpha$ with $i$ numbering the line in the corresponding vacuum diagram. The loop momentum running along the line $i$ in the vacuum diagram is $q_i$. It is a linear combination of the loop momenta. We choose momenta of incoming lines to be $p_{i,\alpha+1}-p_{i\alpha}$ so that the momentum of the line $i\alpha$ is $q_i+p_{i\alpha}$. Concervation of momentum is imposed at the vertices so that $p_{i,d}=p_{j,0}+p_{k,0}$,$q_i=q_j+q_k$ with the chosen orientation of the edges. }
\end{figure}

We adopt the following numbering scheme of lines in the diagram. If we truncate all the external lines, we end up with a vacuum-to-vacuum diagram, which we will call the vacuum diagram corresponding to the diagram in question. We number the lines of this diagram by $i=1,...,I$. The internal lines that constitute the line $i$ in the vacuum diagram will be numbered by an additional index $\alpha$. It was shown in ~\cite{Srednyak} that the general case can be reduced to the case when there are no more than $d+1$ such lines. We will assume this restriction in the following. Therefore, there are the following Feynman denominators 
\begin{equation}
D_{i\alpha}=(q_i+p_{i\alpha})^2+m^2_{i\alpha} , \alpha=0,1,...,\alpha_i , \alpha_i \leq d+1
\end{equation}
that correspond  to the line $i$ in the vacuum diagram. 

We chose a convenient basis for external momenta $p_{i\alpha}$ so that the internal line $(i,\alpha)$ carries momentum $p_{i\alpha}$. Not all of these momenta are independent. There is concervation of momentum relations, of the type 
\begin{equation}
p_{k,0}=p_{i\alpha_i}+p_{j\alpha_j}
\end{equation}
for a triple vertex (for the numbering at Fig.1). In addition, $L$ of them can be set to zero by overall translation invariance. These are all the relations that restrict the possible values of momenta. Therefore, the external momentum space is the affine complex space $\mathbb{C}^N$ of certain dimension ( we treat the generic stratum of paremeter space, and do not worry about compactifying the domain. This question is of central importance for determining global analytic structure and will be treated elsewhere.) 

The momentum $q_i,i=1,...,I$ running along the line $i$ is related to the chosen set of loop momenta $q_a,a=1,...,L$ as follows
\begin{equation}
q_i=l_{ia}q_a
\end{equation}
(summation is assumed over $a$) where $l_{ia}=0,+1,-1$ are the combinatorial coefficients that allow to parametrize the internal line momenta in terms of loop momenta ~\cite{Weinz1}. 


Introduce the following shorthand
\begin{equation}
x_i=\sum_{\alpha=0}^{\alpha_i} x_{i\alpha}.
\end{equation}
The following $L\times L$ matrix
\begin{equation}
A_{ab}=\sum_i l_{ia} l_{ib}x_i
\end{equation}
controls the quadratic form in $q_a$ in the Feynman parametrization. It is convenient to introduce the following vectors 
\begin{equation}
B_i=\sum_{\alpha=0}^{\alpha_i} x_{i\alpha}p_{i\alpha}
\end{equation}
Here we treat the index $i$ as vectorial index in $I$-dimensional space $\mathbb{C}^I$. The vectors $B_i$ have implicit Lorenz index, co that $B_i$ in fact stands for $B_{i\mu}, \mu=1,...,d$. We will suppress this Lorenz index to keep formulas transparent.

We will use also the following vectors 
\begin{equation}
B_a=\sum_i l_{ia} B_i
\end{equation}
Using this notation, the expression for the graph polynomials becomes short
\begin{gather*}
G=det(A)\\
F=-B^T\hat{A}B+G(\sum_{i\alpha} x_{i\alpha} (p^2_{i\alpha}+m_{i\alpha}^2))
\end{gather*}
$\hat{A}$ is the adjoint matrix.


\section{Reduction theorem and its proof}

In this section we state and prove our main theorem. For the background on Schl\"{a}fli function and its various versions see Appendix.

{\bf Theorem 1: } The general Feynman integral $I_\omega(p_{i\alpha})$ can be reduced to a sum of the following integrals
\begin{equation}
J_\beta(p_{i\alpha})=\int_{\bigtriangleup^I} \delta(1-\sum x_i) dx_1\wedge...\wedge dx_I G^v x^\beta Sch(V(x))
\end{equation}
where 
\begin{equation}
Sch(V(x))=Sch_{\Delta(x_1)\times...\times\Delta(x_I)}(\hat{A}_{ij}(x_k)p_{i\alpha}p_{j\beta},G*(p^2_{i\alpha}+m^2_{i\alpha}-m_{i,0}))
\end{equation}

{\bf Proof:}

The proof starts with the following observation. The polynomial $F$ can be rewritten as
\begin{gather*}
F=-\sum_{ij} B_i \hat{A}_{ij}(x_k) B_j+G\sum_{i\alpha} x_{i\alpha}t_{i\alpha}\\
t_{i\alpha}=p^2_{i\alpha}+m^2_{i\alpha}
\end{gather*}
where 
\begin{equation}
\hat{A}_{ij}(x_k)=\sum_{ab} l_{ia} l_{jb} \hat{A}_{ab}(x_k).
\end{equation}
Note that the functions $\hat{A}_{ij}(x_k)$ and $G$ depend only on the sums $x_k=\sum_\alpha x_{i\alpha}$. Therefore we can change integration variables in such a way as to integrate over the $x_{i\alpha}$ first, keeping the sums $x_i$ fixed, and then to integrate over the variables $x_i$. This can be conveniently accomplished by inserting the following identity
\begin{equation}
1=\int dx_1 ...\int dx_I \delta(x_1-\sum_\alpha x_{1\alpha})...\delta(x_I-\sum_\alpha x_{I\alpha})
\end{equation}
Note that the $x_i$ sum to 1. Then the integral transforms into
\begin{gather*}
I_\omega(p)=\int_{\Delta^I(1)} dx_1...dx_I G^v \int_{\Delta^{\alpha_1}(x_1)}...\int_{\Delta^{\alpha_I}(x_I)} \delta(x_1-\sum_\alpha x_{1\alpha})...\\
...\delta(x_I-\sum_\alpha x_{I\alpha}) x^\omega F^w
\end{gather*}
We conveniently rewrite the polynomial $F$ as
\begin{equation}
F=(\sum_{i\alpha} t_{i\alpha}x_{i\alpha})G -\sum_{i\alpha j\beta}\hat{A}_{ij}(x_k) p_{i\alpha}\cdot p_{j\beta} x_{i\alpha}x_{j\beta}
\end{equation}
The inner integral
\begin{equation}
J=\int_{\Delta^{\alpha_1}(x_1)}...\int_{\Delta^{\alpha_I}(x_I)} \delta(x_1-\sum_\alpha x_{1\alpha})...\delta(x_I-\sum_\alpha x_{I\alpha}) x^\omega F^w
\end{equation}
is reducible to a sum of Schl\"{a}fli functions, as we now show. Consider the quadratic term
\begin{equation}
Q(x)=\sum_{i\alpha j\beta}A_{ij}(x_k) p_{i\alpha}p_{j\beta} x_{i\alpha}x_{j\beta}
\end{equation}

Some of the rows of this matix are identically zero. They correspond to the momenta that can be put to zero by overall shift of loop variables. Such $x_{i0}$ can be explicitly integrated out, because $F$ depends linearly on them. This integration modifies $F$ in such a way as to cancel the corresponding rows of the above quadratic form and replace some of the $t_{i\alpha}$ by $t_{i\alpha}-m^2_{i,0}$. We will assume that such modification has been done. 

{\bf Lemma:} The modified quadratic form $Q$ is nondegenerate.

{\bf Proof.}

The proof consists of choosing a spanning tree for the diagram and replacing some of the momenta $p_{i0}$ or $p_{i\alpha_i}$ according to the concervation equations and examining the obtained expression. 

Therefore the integrand is reduced to one of the versions of the generalized Schl\"{a}fli function discussed in the Appendix. It only remains to prove that the domain of integration is reducible to the standard simplex. This is established in the following 

{\bf Lemma 2:} The product of simplices $\Delta^\alpha_1\times...\Delta^\alpha_I$ can be tringulated by standard simplices.

{\bf Proof:}This follows from more general results ~\cite{Ziegler1}. 

We will write the triangulation map as
\begin{equation}
\Delta^\alpha_1(x_1)\times...\times \Delta^\alpha_I(x_I) \rightarrow \cup_r \Delta^N_r
\end{equation}
Each of the elementary simplices in the union has dimension $(\alpha_1-1)+...+(\alpha_I-1)$.

The proof of the theorem can be finished by observing that the integral with the factor $x^\omega$ can be represented by a sum of Schalfli functions with algebraic coefficients, as is proved in the Appendix. This finishes the proof of the theorem.

{\bf Corollary } It is possible to reduce the inner Schl\"{a}fli integral to the following form
\begin{equation}
J'=\int_{\Delta(V(x))} (1-y_1^2-...-y_N^2)^w dy_1...dy_N
\end{equation}
for certain $x_i$-dependent simplex $\Delta=\Delta(x_k)$.

The simplex in this corollary is obtained by a choice of $N+1$ points among the images of the vertices of the polyhedron $\Delta_1\times...\times\Delta_I$ under a linear transformation
\begin{equation}
(x_{i\alpha}) \rightarrow y_{i\alpha}=S_{i\alpha j_\beta} x_{j\beta}+c_{i\alpha}
\end{equation}
that diagonalizes the matix 
\begin{equation}
V_{i\alpha j\beta}=\hat{A}_{ij} p_{i\alpha}p_{j\beta}
\end{equation}


\section{Examples}

{\bf 1-loop}

The theorem is vacuous in this case. Indeed, in Feynman parametrization in the example of 4d scalar theory 1-loop diagram has the form
\begin{equation}
I=\int_{\Delta} \delta(1-\sum x_i) (\sum x_i(p_i^2+m_i^2)-(\sum x_i p_i)(\sum x_j p_j))^{-1}
\end{equation}
This is a Schl\"{a}fli function from the start.
\newpage

{\bf 2-loop}

The diagram in this case is drawn in Fig. 2. We consider 4-dimensional scalar field theory. The integral in question is given by
\begin{gather*}
I=\int 1/\prod_{\alpha=0}^4 ((q_1+p_{1\alpha})^2+m_{1\alpha}^2)\prod_{\alpha=0}^4 ((q_2+p_{2\alpha})^2+m^2_{2\alpha}) \\
\prod_{\alpha=0}^3 ((q_1+q_2+p_{3\alpha})^2+m^2_{3\alpha})((q_1+q_2+P_{12})^2+m^2_{3,4})
\end{gather*}
We denoted $P_{12}=p_{1,4}+p_{2,4}$ and accounted for the conservation of momentum. Note that the product corresponding to the line carrying loop momentum $q_1+q_2$ has only 4 terms. 

\begin{figure}
\centering\includegraphics[width=0.5\textwidth]{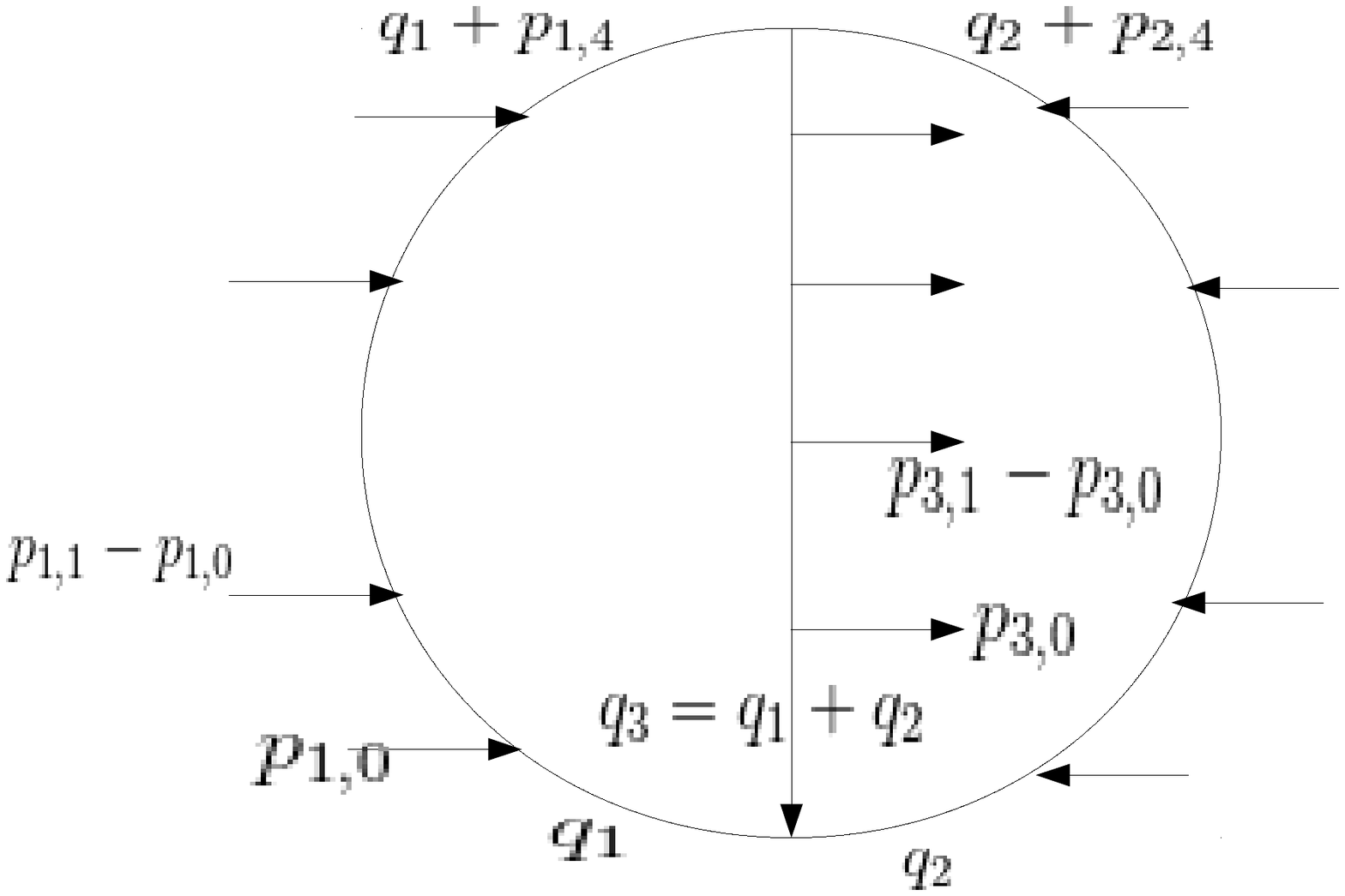}
\caption{The generic 2-loop diagram. The convention on the external momenta is such that each line carries the momentum $q_i+p_{i\alpha}$. Concervation of momentum relations are $p_{3,4}=p_{1,4}+p{2,4}$,$q_3=q_1+q_2$. }
\end{figure}

After the introduction of Feynman parameters we get the following polynomial
\begin{multline}
\sum x_{i\alpha} D_{i\alpha}=\sum_\alpha x_{1\alpha} ((q_1+p_{1\alpha})^2+m^2_{1\alpha})+\\
+\sum_\alpha x_{2\alpha} ((q_2+p_{2\alpha})^2+m^2_{2\alpha})+\\
+\sum_\alpha x_{3\alpha} ((q_1+q_2+p_{3\alpha})^2+m^2_{3\alpha})+\\
+((q_1+q_2+P_{1,2}^2+m^2_{3,4})(1-\sum_{i\alpha} x_{i\alpha})
\end{multline}

After some algebra we obtain the following expressions
\begin{equation}
A = \begin{bmatrix}
       1 & 1-\sum_{\alpha=0}^4 x_{1\alpha}-\sum_{\alpha=0}^4 x_{2\alpha} \\
       1-\sum_{\alpha=0}^4 x_{1\alpha}-\sum_{\alpha=0}^4 x_{2\alpha}  & 1 \\
     \end{bmatrix}
\end{equation}
\begin{equation}
\hat{A} = \begin{bmatrix}
       1 & -(1-\sum_{\alpha=1}^4 x_{1\alpha}-\sum_{\alpha=1}^4 x_{2\alpha}) \\
       -(1-\sum_{\alpha=1}^4 x_{1\alpha}-\sum_{\alpha=1}^4 x_{2\alpha})  & 1 \\
     \end{bmatrix}
\end{equation}

The graph polynomial after $q$ integration takes form
\begin{multline}
F=
-\begin{bmatrix}
	p_1  \\
	p_2 \\
	p_3\\
\end{bmatrix}^T
\begin{bmatrix}
	\hat{A}_{11}  & \hat{A}_{11} & \hat{A}_{11}+\hat{A}_{11} \\
	\hat{A}_{11} & \hat{A}_{11} & \hat{A}_{11}+\hat{A}_{11} \\
	\hat{A}_{11}+ \hat{A}_{11} & \hat{A}_{11}+\hat{A}_{11} & \hat{A}_{11}+2\hat{A}_{11}+\hat{A}_{11} 
\end{bmatrix}
\begin{bmatrix}
	p_1  \\
	p_2 \\
	p_3\\
\end{bmatrix}-\\
\sum (x_{i\alpha}(p^2_{i\alpha}+m^2_{i\alpha})det(A)
\end{multline}
where we denoted 
\begin{equation}
p_i=\sum_\alpha x_{i\alpha} p_{i\alpha}
\end{equation}

We can choose 
\begin{equation}
p_{1,0}=0, p_{2,0}=0
\end{equation}
due to overall loop momentum shift invariance. Concervation of momentum imposes
\begin{equation}
p_{3,4}=p_{1,4}+p_{2,4}
\end{equation}

We can perform the integration over $x_{1,0}, x_{2,0},x_{3,0}$ because the matrix does not depend on these variables. The result of this integration corresponds to replacing 
\begin{equation}
p^2_{i\alpha}+m^2_{i\alpha} \rightarrow p^2_{i\alpha}+m^2_{i\alpha} -M^2
\end{equation}
with $M=m_{1,0},m_{2,0},m_{3,0},0$. 

Therefore the intergal is reduced to the sum of integrals of the following type
\begin{equation}
J=\int_{\Delta^3}\delta(1-\sum x_i) Sch^w_{\Delta^4(x_1)\times\Delta^4(x_1)\times\Delta^4(x_3)}(A_{ij}p_{i\alpha}p_{j\beta}, det(A)t_{i\alpha},C)
\end{equation}
The Schl\"{a}fli function can be algebraically expressed through the standard generalized Schl\"{a}fli function of dimension $11$, according to the Appendix. There remain $2$ integrations over the variables $x_i$.


\section{Conclusion}

We demonstrated that most of the integrations in a Feynman integral with many external lines can be performed in terms of a family of functions that have nice analytic, geometric and number theoretic properties - the Schl\"{a}fli functions. A large body of knowledge about the properties of these functions was accumulated in the mathematics literature. We wish to conclude with posing the questions by answering which the contents of this paper can be made more trnasparent and precise.

{\bf Q1: } We proved that the Schl\"{a}fli function $Sch_{\Delta_1\times...\times\Delta_n}$ can be algebraically related to the standard inhomogeneous Schl\"{a}fli function. This reduction is nontrivial. It involves relative positions of the vertices of the elementary simplices in the spaces $(x_{i\alpha})$ after transformation by the matrix that diagonalizes the momentum matrix 
\begin{equation}
V=A_{ij}p_{i\alpha}p_{j\beta}
\end{equation}
This matrix is block diagonal. The proper understanding of this reduction should involve flag varieties and parabolic groups. It will also shed new light on the geometry of multiparticle configuration spaces in perturbation theory.

{\bf Q2:  } It is shown by ~\cite{Ao1} that conformal Schl\"{a}fli functions are representable by iterated integrals. Since then the theory of polylogarithms has developed greatly. What is the precise relation between the Schl\"{a}fli function and polylogarithms? 

{\bf Q3: } The equations defining the singular loci and the way Schl\"{a}fli function branches along them, which implies that it lifts to multiple branched covers, suggests a re-interpretation in terms of twistor geometry and its generalizations. What is this relationship, precisely?



\section{Appendix: Some properties of generalized Schl\"{a}fli function}

There is a number of similar functions associated to intergals of a quadratic form over a simplex or product of standard simplices. We write them out in turn and show that they all are algebraically related to the standard inhomogeneous generalized Schl\"{a}fli function. We start with symmetric homogeneous Schl\"{a}fli function
\begin{equation}
I(V)=\int_\Delta \delta(1-\sum x_i) (\sum_{ij}V_{ij}x_i x_j)^w d^nx
\end{equation}
or its inhomoheneous version
\begin{equation}
I(V_{ij},V_i,V_0)=\int_\Delta (\sum_{ij}V_{ij}x_i x_j+\sum_i 2V_ix_i+V_0)^w d^nx
\end{equation}

The integration is performed over the standard simplex $\Delta$, which has $n+1$ boundary components that lie in coodinate hyperplanes. The integral makes sense for complex matrices, making it a function on an open subset of the space of $(n+1)\times(n+1)$ complex matrices, assembled from the coefficients of the linear forms $f_i$. We refer to this function as the Schl\"{a}fli function in dimension $n$.

The following function was considered in ~\cite{Ao1}
\begin{equation}
J(f_0,...,f_n)=\int_{\Delta(f_0,...,f_n)} (1-x_1^2-...-x_n^2)^w dx_1...dx_n
\end{equation}
 The simplex of integration ${\Delta(f_0,...,f_n)}$ was defined by the conditions $f_i(x)=f_{i0}x_0+...+f_{in}x_{in}x_n\geq 0$ for real $f_{ij}$ and then analytically continued in the complex domain. 

{\bf Lemma: } The function $J$ can be transformed to inhomogeneous Schl\"{a}fli function $J$. 

{\bf Proof:} It is enough to observe that by a linear transformation of variables 
\begin{equation}
x_i=\sum_j T_{ij}y_j+a_i
\end{equation}
it is possible to map the simplex $\Delta(f_0,...,f_n)$ to the standard simplex. This proves the lemma.

The most convenient function for the expression of Feynman integrals is the following
\begin{gather*}
Sch^w_{\Delta^{\alpha_1}(x_1)\times...\Delta^{\alpha_n}(x_n)}(V_{i\alpha j\beta},V_{i\alpha},V_0)=
\int_{\Delta^{\alpha_1}(x_1)}(dx_{1\alpha})\delta(x_1-\sum_\alpha x_{1\alpha})...\\
\int_{\Delta^{\alpha_n}(x_n)}(dx_{n\alpha})\delta(x_n-\sum_\alpha x_{n\alpha})
(\sum V_{i\alpha j\beta} x_{i\alpha}x_{j\beta}+\sum V_{i\alpha} x_{i\alpha}+V_0)^w
\end{gather*}

{\bf Lemma:} The function $Sch_{\Delta^{\alpha_1}(x_1)\times...\Delta^{\alpha_n}(x_n)}(V_{i\alpha j\beta},V_{i\alpha},V_0)$ 
is reducible to a sum of inhomogeneous Schl\"{a}fli functions $J$ of dimension $(\alpha_1-1)+...+(\alpha_n-1)$.

{\bf Proof:} It is enough to observe that the $\delta$-functions can be used to reduce the number of integrations to $(\alpha_1-1)+...+(\alpha_n-1)$, and that the product of simplices can be triangulated. The resulting simplices can be normalized to the standard simplex by a change in the matrix coordinates. 

Much more is known about the analytic and number theoretic properties of the Schl\"{a}fli function (see ~\cite{Ao3} and references therein).  We cite just one of them. In ~\cite{Ao1} the following iterated integral representaion was derived in conformal case. The function is written as
\begin{equation}
I(f)=\int_\Delta (1-x_1^2-x_2^2-...-x_n^2)^{1/2}dx_1...dx_n
\end{equation}
where the domain is defined by $f_i(x)\geq 0$. For simplicity, we assumed that the domain projects 1-1 to the coordinate hyperplane $x_1,...,x_n$, otherwise it is necessary to use the Euler form. 

Denote by $a_{ij}$ the cosine of the angle between the hyperplanes $f_i(x)=0$ and $f_j(x)=0$ (continued in the complex domain). Denote by $D(\frac{I}{J}), I=(i_1,...,i_p),J=(j_1,...,j_p)$ the $I,J$-th minir of the following matrix
\begin{equation}
A=\begin{bmatrix}
1 & a_{12} & \dots a_{1,n+1} \\
a_{21} & 1 & \dots a_{2,n+1} \\
\dots \\
a_{n+1,1} & a_{n+1,2}&  \dots 1 \\
\end{bmatrix}
\end{equation}
Denote $D(I,I)$ by $D(I)$ and introduce the following 1-forms
\begin{multline}
\omega(I,J)=-i/2\times \\
dln\frac{-D(\frac{(i_1,i_2,...,i_p,i_{p+1})}{(i_1,i_2,...,i_p,i_{p+2})})+i\sqrt{D((i_1,i_2,...,i_p)) D((i_1,i_2,...,i_{p+1},i_{p+2}))} }{ -D(\frac{(i_1,i_2,...,i_p,i_{p+1})}{(i_1,i_2,...,i_p,i_{p+2})})-i\sqrt{D((i_1,i_2,...,i_p))) D((i_1,i_2,...,i_{p+1},i_{p+2}))}}
\end{multline}

Then for n odd $n=2\nu+1$ the Schl\"{a}fli function is represented by iterated integral 
\begin{equation}
I(f)=\sum_{I_0,I_2,...,I_\nu} \sum_{\sigma=0}^\nu \int_E^A \omega(I_0,I_1)\omega(I_1,I_2)...\omega(I_{\sigma-1},I_\sigma)\frac{|S^{n-2\sigma}|}{2^{n+1-2\sigma}}
\end{equation}
where the sum is extended over increasing chains of indices (this is Theorem 1 of ~\cite{Ao1}). 

Explicit form of the Gauss-Manin connection involves many nontrivial terms. From the above analysis we see that there are an order $2^n$ components in the branching divisor. Therefore the rank of the bundel defined by the integral is of order $2^n$. The explicit form of the connection will be addressed in subsequent publications.



\bibliographystyle{amsplain}
\bibliography{sample}

\end{document}